# Improved scintillation proportionality and energy resolution of LaBr$_3$:Ce at 80K

Ivan V. Khodyuk, Mikhail S. Alekhin, Johan T.M. de Haas, and Pieter Dorenbos
Luminescence Materials Research Group, Faculty of Applied Sciences, Delft University of Technology, Mekelweg 15, 2629 JB Delft, the Netherlands

**Abstract**

Using highly monochromatic synchrotron X-rays in the energy range from 10.5 keV to 100 keV the temperature dependence of nonproportionality and energy resolution of LaBr$_3$ scintillators doped with 5% Ce$^{3+}$ were studied at 80K, 295K, and 450K. Improvement of the proportionality and better energy resolution was observed on lowering the temperature. This effect suggests that the already outstanding energy resolution of LaBr$_3$:Ce can be improved even further. It also may provide new clues to better understand the processes that cause nonproportionality of inorganic scintillator response.

## 1. Introduction

The widespread use of inorganic scintillators for applications in science and society is the driving force behind the search for new high performance compounds. Both novel and known phosphors are constantly suggested as promising scintillators [1-4]. The most important requirement imposed on new scintillators is a high energy resolution ($R$) for gamma ray detection. $R$ is defined as the Full Width at Half Maximum (FWHM) $\Delta E$ over the energy $E$ of the full absorption peak in so-called pulse height spectra [5]. There are two key factors that determine the energy resolution, i.e., Poisson statistics in the number of detected photons ($R_M$) and the nonproportionality of the light yield of scintillators with gamma-ray energy ($R_{nPR}$) [6]:

$$\left(\frac{\Delta E}{E}\right)^2 = R^2 = R_M^2 + R_{nPR}^2; \qquad (1)$$

$$R_M = 2.35\sqrt{(1+\mathrm{var}(M))/N_{phe}^{PMT}}, \qquad (2)$$

where var$(M)$=0.28 is the contribution from the fractional variance in the gain of the Hamamatsu R6231-100 photomultiplier tube [7] (PMT) and $N_{phe}^{PMT}$ is the number of photoelectrons produced in the PMT by the scintillation photons from LaBr$_3$:Ce.

Nonproportionality means that the total light output of a scintillator is not precisely proportional to the energy of the absorbed gamma-ray photon. This has a deteriorating effect on energy resolution [6]. For example, based on Poisson statistics alone LaBr$_3$:Ce, which was discovered in 2001 [8], should display an energy resolution of 2.1% when 662 keV gamma-ray photons are detected with a standard type PMT. However in reality it is about 2.8%. Because the light yield and the PMT performance is already close to optimal we need to reduce $R_{nPR}$ in order to improve the energy resolution, and for that we wish to understand the physical causes of nonproportionality.

Nonproportionality is due to electron-hole recombination losses during the scintillation process [9]. It is currently believed that those losses occur inside parts of the ionization track where the ionization density is high [10-12]. That density increases when the gamma-ray energy decreases. The scintillation yield per energy unit in LaBr$_3$:Ce scintillator at 10 keV energy is for example 15% smaller than at 662 keV [5]. The origin of this decrease in efficiency, i.e., the true cause of electron-hole recombination losses, and the related deterioration in energy resolution is not known. It is a mystery to the scintillation community why some scintillators reveal poor proportionality while others appear reasonably good [13]. To elucidate the true origin of nonproportionality, accurate data on the scintillation light output of LaBr$_3$ doped 5% Ce$^{3+}$ as function of energy and temperature were measured.

Attempts to measure the nonproportional response (nPR) as a function of temperature were made before [14] by changing the temperature of both the scintillator and PMT over a relatively narrow temperature range from -30ºC to +60ºC. In our experiments the PMT remains at room temperature while the scintillator temperature can be changed. Thus, we do not need to consider the effect of temperature on the characteristics of the PMT, and we measure the intrinsic properties of the scintillator only.

**2. Experimental**

To measure the pulse height spectra at many finely spaced energy values between 10.5 keV and 100 keV, experiments were carried out at the X-1 beam line at the Hamburger Synhrotronstrahlungslabor (HASYLAB) synchrotron radiation facility in Hamburg, Germany. A highly monochromatic pencil x-ray beam in the energy range 10.5 – 100 keV was used as an excitation source. A description of the experimental setup used for nonproportionality and energy resolution measurements can be found in [15, 16]. To record scintillation pulse height spectra as a function of temperature, a LaBr$_3$:Ce sample was fixed at the bottom of a parabolic-like stainless steel cup covered with reflective Al foil, mounted onto the cold finger of a liquid nitrogen bath cryostat. The cup directed the scintillation light through the quartz window towards a PMT situated outside the cryostat chamber. The Hamamatsu R6231-100 PMT at -680V remained at room temperature and observed about 20% of the emitted scintillation light. To collect as much of the PMT charge pulse as possible, the shaping time of an Ortec 672 spectroscopic amplifier was set at 10 μs. The temperature of the crystal was controlled by two thermocouples attached to different parts of the sample holder.

**3. Results and discussion**

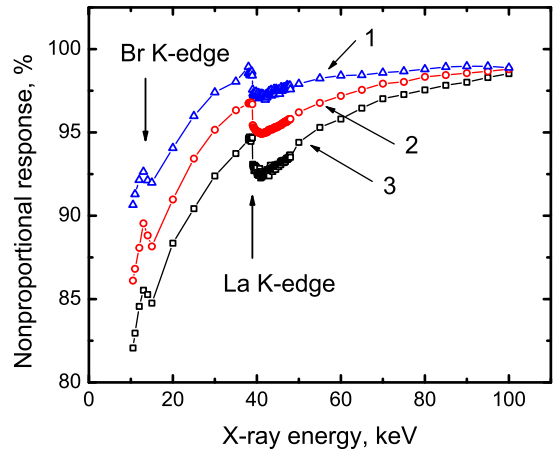

**Figure 1**. Nonproportional response of LaBr$_3$:Ce as a function of x-ray energy (E$_X$) at curve 1) 80K, 2) 295K, and 3) 450K.

We define the nonproportional response (nPR) of a scintillator at x-ray energy (E$_X$) as the light output per MeV observed at energy E$_X$ divided by the light output per MeV observed at E$_X$ = 662 keV. Figure 1 shows the nPR as a function of E$_X$ at 80K, 295K and 450K. The shape of the response curves has been discussed by us in detail previously [5]. Note that for LaBr$_3$:Ce in Fig.1 we observe discontinuities in the nPR curves not only at the lanthanum K-electron binding energy $E_{KLa}$ = 38.925 keV, but also at the bromine K-electron binding energy $E_{KBr}$ = 13.474 keV. But most important observation is that the nPR reveals a strong temperature dependence. At 100 keV the values of nPR at all three temperatures are almost the same 98.8%. The situation is quite different at low energy. At 10.5 keV the nPR at 80 K is 90.7%, at 295 K it falls to 86.1% and at 450 K it has decreased to 82.1%. Therefore, a clear improvement in nPR occurs as the temperature decreases from 450K to 80K. Since the typical error in nPR does not exceed 0.1% at 10.5 keV these are significant improvements. This result indicates that electron-hole recombination losses in the high ionization density part of the ionization track in LaBr$_3$:Ce have a strong temperature dependence.

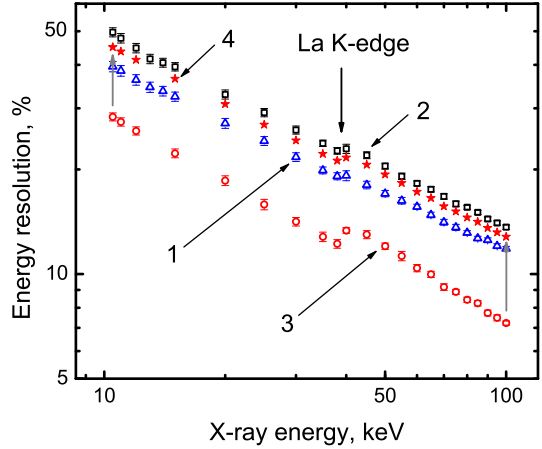

**Figure 2**. Energy resolution of the x-ray photopeak as a function of x-ray energy for LaBr$_3$:Ce inside the cryostat at 80K (curve 1) and at 450K (curve 2). Curve 3) is obtained with LaBr$_3$:Ce mounted on the PMT at 295K. Curve 4), is the resolution derived from 3) for LaBr$_3$:Ce inside the cryostat at 295K.

The resolution $R$ of the X-ray photopeaks is plotted on a double-log scale in Fig. 2 as a function of $E_X$. Ideally when only $R_M$ contributes to the energy resolution a straight line with slope -0.5 is expected, see Eq. (2) and [17]. Due to a relatively small contribution of $R_{nPR}$ to $R$, a slight deviation of the experimental values from a straight line can be observed as a step-like increase of $R$ at energy $E_{KLa}$. This situation was discussed in detail by us in [5]. Curve 1) and 2) in Fig. 2 show data obtained at 80K and 450K with the LaBr$_3$:Ce crystal mounted inside the cryostat. For these data, improvement of energy resolution is clearly seen with decrease of the scintillation crystal temperature. At 100 keV, the energy resolution at 80 K is 1.8% smaller than at 450 K. With a decrease of X-ray energy the resolution difference increases reaching 10% at 10.5 keV. The $N_{phe}^{PMT}$ produced in the PMT by LaBr$_3$:Ce 5% mounted inside the cryostat with 100 keV X-rays at 80 K was 479 photoelectrons while it was 463 photoelectrons at 450 K. This confirms the high temperature stability of the scintillation light yield of LaBr$_3$:Ce 5% [18]. The improved energy resolution at 80 K is therefore attributed to better proportionality.

To further demonstrate that the improved energy resolution is due to improved proportionality we collected data obtained at room temperature by direct optical coupling of the LaBr$_3$:Ce 5% scintillation crystal to the photocathode of a Hamamatsu R6231-100 PMT, see curve 3) in Fig. 2. The number of photoelectrons $N_{phe}^{PMT}$ produced in the PMT by 100 keV x-ray detection was 2254 photoelectrons whereas $N_{phe}^{PMT}$ = 485 when the same crystal is mounted inside the cryostat. Together with the data in Fig.1, Eq. (2), and Eq. (3) this provides us with $N_{phe}^{PMT}(E_X)$, $R_M(E_X)$, and $R_{nPR}(E_X)$ at 295K. Assuming that $R_{nPR}(E_X)$ is an intrinsic contribution from the scintillator that is not affected by the photon collection efficiency we can calculate $R(E_X)$ at room temperature for a scintillator mounted inside the cryostat, see curve 4) in Fig. 2. Note that curve 4) falls nicely in between curve 1) and 3). The increasingly better proportionality observed in Fig. 1 when temperature decreases is therefore fully consistent with the improvement of energy resolution observed in Figure 2.

**4. Conclusions**

Summarizing, we have found that by reducing the temperature of $LaBr_3:Ce^{3+}$ scintillator from 450 K to 295 K to 80K its proportionality improves and as a consequence a better energy resolution for X-ray photon detection is obtained. This improvement evidences that the yet unknown parameters responsible for the nonproportional response of $LaBr_3:Ce$ depend on temperature. Apparently, with decreasing temperature, the amount of electron hole recombination losses in the dense parts of the ionization track decreases and as a consequence the efficiency of the scintillator increases. Recent simulation data [11] suggest that electron and hole mobilities are important parameters that determine nonproportionality, and possibly our finding is a manifestation of changing mobilities with changing temperature. It is known that electron mobility is strongly temperature dependant due to phonon scattering [19]. Although, the cause of proportionality improvement is not yet established in this work, our results do demonstrate that nonproportionality is a property that can be improved. If it can be improved by changing temperature one may hope to improve it also by other means like activator concentration, crystal quality, co-dopants etc.

**Acknowledgment**

This work was funded by the Dutch Technology Foundation (STW), and supported by Saint-Gobain Crystals, France and by the European Community's Seventh Framework Programme (FP7/2007-2013) under grant agreement n° 226716. We thank the scientists and technicians of the X-1 beamline at the Hamburger Synhrotronstrahlungslabor (HASY-LAB) synchrotron radiation facilities for their assistance.